\documentstyle[11pt]{article}
\input amssym.tex
\begin{document}
\newcommand{\be}{\begin{equation}}
\newcommand{\bse}{\begin{subequations}}
\newcommand{\ese}{\end{subequations}}
\newcommand{\bea}{\begin{eqnarray}}
\newcommand{\eea}{\end{eqnarray}}
\newcommand{\ba}{\begin{array}}
\newcommand{\ea}{\end{array}}
\newcommand{\ee}{\end{equation}}
\newcommand{\vm}{\vspace{0.2cm}}
\newcommand{\vl}{\vspace{0.4cm}}
\renewcommand{\theequation}{\thesection.\arabic{equation}}
\begin{flushright}
HIP-2005-56/TH
\end{flushright}

\vspace{0.2cm}
\begin{center}
\begin{Large}
{\Large\bf Connection between Color Confinement\\
\vskip 5mm and the Gluon Mass}
\end{Large}
\vskip 10mm {{\bf Masud Chaichian}}

\vspace{0.2cm}
      {\small   Department of Physical Sciences, High Energy Physics Division}\\
           {\small University of Helsinki} \\
{\small and}\\ {\small Helsinki Institute of Physics}\\ {\small P.O.
Box 64, FIN-00014 Helsinki, Finland}\\
and\\\vskip 2mm{\bf Kazuhiko Nishijima}

\vspace{0.2cm}
 {\small
Department of Physics, University of Tokyo, 7-3-1 Hongo,\\
Bunkyo-ku, Tokyo 113-0033, Japan}
\date{}
\end{center}


\vspace{2.0cm}
\begin{center}\begin{minipage}{5in}
\begin{center} {\bf Abstract} \end{center}
\baselineskip 0.3in{\small It is argued that there are two phases in
QCD distinguished by different choices of the gauge parameter. In
one phase the color confinement is realized and gluons turn out to
be massive, whereas in the other phase it ceases to be realized, but
the gluons remain massless.
}\\

\vskip 10mm
\end{minipage}
\end{center}

\newpage

\begin{center}
{\Large\bf Foreword}
\end{center}

It is a great honor for us to take part in the celebration of our
dear friend, Professor Adriano Di Giacomo, on the occasion of his
seventieth anniversary, and we feel a sense of privilege in
dedicating this article to him. Adriano and we share a common
interest in understanding color confinement (see, e.g.
\cite{DG1,DG2} and references therein) so that we shall concentrate
our attention to the exploration of the connection between
confinement and the gluon mass in this article.

\section{Quest for Massless Particles}

Already in classical physics we know two kinds of fundamental
interactions, namely, Coulomb and gravitational interactions. They
have been recognized among many other phenomenological interactions
by their long-range character obeying the inverse square law. In
this connection it is interesting to observe that these two kinds of
forces dominate in regions of different scales. At astronomical
distances the gravitational forces dominate, but at microscopic
distances the Coulomb forces take over. For instance, between two
protons the gravitational force is weaker than the Coulomb force by
36 orders of magnitude. The rather vague border between these
regimes is defined as distances at which the gravitational and
electromagnetic forces balance. At the time of Millikan the size of
an oil drop represented the border, but with the rapid progress of
technology even the size of a linear motor car can be identified as
a contemporary border.

Then a question is raised of why they dominate at different scales.
In answering this question we may attribute the cause to the
difference in the tensorial ranks of the fields mediating these
interactions. The gravitational interactions are mediated by a
second rank tensor field, whereas the electromagnetic interactions
are mediated by a vector field.

In the former case both particle-particle and particle-antiparticle
interactions are attractive, leading to the well-known universal
attraction. In the latter case, however, only particle-antiparticle
interactions are attractive, while particle-particle interactions
are repulsive. It seems to be likely that there are no other forces
obeying the inverse square law than those mentioned above. Thus in a
macroscopic system like charges repel each other since there are no
stronger attractive forces at such a scale and macroscopic systems
tend to be electrically neutral. In a microscopic system, however,
this is not the case since there are stronger attractive nuclear
forces that overcome the Coulomb repulsion, and consequently we find
multiply-charged nuclei. In this way, as the size of a system
increases, the electromagnetic forces turn out to be less important
and start to be taken over by gravity.

In quantum field theory, long-range forces obeying the inverse
square law are generated by fields of massless quanta. Then it
occurs to us that there would be no massless fields other than those
mentioned above, since there is no experimental evidence for such a
field in the macroscopic world. In the microscopic world, however,
the gluon - the quantum of the color gauge field - is a possible
candidate for a massless particle, and we shall check this
possibility in what follows.

The interaction between two electrically neutral systems is given by
the van der Waals potential, proportional to $r^{-6}$. This shows
that the electric fields generated by neutral systems can penetrate
into the vacuum without any cut-off. This should be compared with
the short-range character of strong interactions that are typically
represented by the Yukawa potential between hadrons. Hadrons are
color singlets and the color gauge field generated by them cannot
penetrate into the vacuum beyond a certain limit of the order of the
pion Compton wave length. Otherwise, the strong interactions would
dominate the electromagnetic ones at all scales, in contradiction to
our experience.

The above observation indicates that the gluons are confined within
hadrons. Then it reminds us of the proton-electron model of the
nucleus in the old days, in that it was very difficult in this model
to confine light electrons inside a nucleus, since the uncertainty
relation implies very high momentum, of the order of 100 MeV, for
the electrons. In the case of strong interactions we are confronted
with a similar difficulty again, since the uncertainty relation
implies very high kinetic energies for the massless gluons, so that
it seems very difficult, if not impossible, to confine them within a
tiny hadron. A plausible way out of this difficulty would be to
assume reasonably massive gluons, so that their kinetic energies
turn out to be sufficiently low so as not to leak out of the hadron.
On the other hand, when gluons are not confined there is no reason
to believe why they should be massive.

Thus we predict a possibility of the presence of a connection
between color confinement and the gluon mass and in the following
sections we shall justify this intuitive reasoning theoretically
\cite{color_conf}.

\setcounter{equation}{0}
\section{Condition for Color Confinement}
The problem of color confinement has been discussed elsewhere in
detail \cite{color_conf}-\cite{4} and we shall briefly state the
condition for color confinement in our version. For this purpose we
are going to introduce some notation, and start from the Lagrangian
density of QCD:
\be
{\cal L} = {\cal L}_{inv}+ {\cal L}_{gf}+{\cal L}_{FP}, \ee
where
\bea\label{lagr_qcd}{\cal L}_{inv}&=& -\frac{1}{4}F^{\mu\nu}\cdot
F_{\mu\nu} +i \overline{\psi}(\gamma^{\mu} D_{\mu}-m)\psi\ ,\cr
{\cal L}_{gf}&=&-A^{\mu}\cdot
\partial_{\mu} B +\frac{\alpha}{2} B\cdot B\ ,\cr{\cal L}_{FP}&=& -i\partial^{\mu}
\overline{c} \cdot D_{\mu}c  \eea
in the customary notation. We have suppressed the color and flavor
indices in (\ref{lagr_qcd}).  The second term ${\cal L}_{gf}$ is the
gauge-fixing term in which
 $\alpha$ denotes the gauge parameter and
$B$ the Nakanishi-Lautrup auxiliary field. The last term ${\cal
L}_{FP}$ is the Faddeev-Popov (FP) ghost  term, and the
anticommuting scalar fields $c$ and $\bar{c}$ denote the FP ghost
fields. Only the first term ${\cal L}_{inv}$ is invariant under
local gauge transformations, but the total Lagrangian is invariant
under the global Becchi-Rouet-Stora (BRS) transformations
\cite{BRS}.

Let us consider an infinitesimal gauge transformation of the gauge
and quark fields and replace the gauge function either by $c$ or by
$\bar c$. Then they define two kinds of BRS transformations denoted
by $\delta$ and $\bar \delta$, respectively:
\bea \delta A_{\mu}&=& D_{\mu}c ,\ \   \overline{\delta} A_{\mu}=
D_{\mu}\overline{c} ,\\
 \delta \psi&=& ig (c\cdot T) \psi ,\ \
\overline{\delta}  \psi= ig (\overline{c}\cdot T) \psi . \eea
where the matrix $T$ is introduced in the covariant derivative of
$\psi$ as
\be D_{\mu} \psi = (\partial _{\mu} -i g T \cdot A_{\mu} ) \psi\ ,
\ee
For the auxiliary fields $B$, $c$ and $\bar{c}$ the BRS
transformations are determined by requiring the invariance of the
local Lagrangian density, namely,
\be \delta {\cal L} = \overline{\delta} {\cal L}= 0\ . \ee
We shall not write them down explicitly since they are not relevant
to the following arguments.

Then the equations of motion for  the gauge field can be expressed
with the help of BRS transformations as
\be\label{eq-motion}
\partial^{\mu}F_{\mu\nu} + gJ_{\nu}= i\delta\overline{\delta}
 A_{\nu}\ ,
\ee
where $J_{\nu}$ denotes the color current density and $g$ the gauge
coupling constant. It is worth noting that all three terms in
(\ref{eq-motion}) are divergenceless separately, in particular
\be
\partial^{\nu}(i\delta\overline{\delta}
 A_{\nu}) =0\ .
\ee
The generator of the BRS transformation $\delta$ is denoted by
$Q_B$, then the physical states are defined as those states that are
annihilated by applying $Q_B$. In this way we can safely eliminate
the unphysical states of the indefinite metric. We interpret color
confinement that colored particle states are unoservable, since they
are not physical in the above sense. Then, as has been discussed
before \cite{color_conf}-\cite{4}, the condition for color
confinement is simply expressed by
\be\label{conf_cond} C=0\ , \ee
where the constant $C$ is defined by
\be
\partial^{\nu} \langle i \delta \bar{\delta} A_{\nu}^{a} (x),
A_{\sigma}^{b} (y) \rangle = i \delta_{ab} C \partial _{\sigma}
\delta ^4 (x-y)\ . \ee
Here $a$ and $b$ are color indices and $\langle\cdots\rangle$
denotes the vacuum expectation value of the time-ordered product.
Furthermore, with the help of the renormalization group (RG) and
asymptotic freedom \cite{Gross,Politzer}, it has been shown that the
conditions (\ref{conf_cond}) is equivalent to
\be Z_3^{-1}=0 , \ee
where $Z_3$ is the renormalization constant of the color gauge
field. We shall come back to this subject later.

\setcounter{equation}{0}
\section{Connection between  Color Confinement and the Gluon Mass}

With the help of (\ref{eq-motion}) we can write down an equation for
two-point Green's functions of the form
\be\label{green_funct} \langle
\partial^{\lambda}F^a_{\lambda\mu}(x),A_{\nu}^b(y)\rangle +
\langle g{J}^a_{\mu}(x),A_{\nu}^b(y)\rangle =\langle
i\delta\overline{\delta}A_{\mu}^a ,A_{\nu}^b(y)\rangle\ . \ee
We shall study the structure of the Fourier transforms of these
Green's functions.

Let $F_{\mu}$ and $G_{\nu}$ be vector fields and introduce
\be \langle F_{\mu}(x),G_{\nu}(y)\rangle =\frac{-i}{(2\pi)^4}\int
d^4k e^{ik\cdot(x-y)} T_{\mu\nu}(k) , \ee
and the Fourier transform of $\langle F_{\mu},G_{\nu}\rangle$ is
denoted by
\be T_{\mu\nu}(k) =\mbox{FT}\langle F_{\mu},G_{\nu}\rangle . \ee
Then $T_{\mu\nu}$ can be expressed as a linear combination of two
covariants:
\be T_{\mu\nu}(k)=- \frac{k_{\mu}k_{\nu}}{k^2+i\epsilon}T_0(k^2)
-\left(g_{\mu\nu} - \frac{k_{\mu} k_{\nu}}{k^2+i\epsilon}\right)
T_1(k^2) . \ee

Next we introduce two conditions:

1) Assume $\partial^{\mu}F_{\mu}=0$, then we have
\be T_0(k^2)=T_0, \ \mbox{constant}. \ee

2) Assume further that $Q\vert0\rangle=\langle 0\vert Q=0$, where
\be Q=\int d^3x\ F_0(x), \ee
then we  have
\be T_0=0 . \ee
All three terms in (\ref{green_funct}) satisfy the constraint 1) and
their Lehmann representations are given as follows. First, by taking
account of the antisymmetry between subscripts $\lambda$ and $\mu$,
we find
\be \mbox{FT}\langle F_{\lambda\mu},A_{\nu}\rangle
=-i(k_{\lambda}g_{\mu\nu} - k_{\mu}g_{\lambda\nu})\left[
\frac{R}{k^2+i\epsilon} + \int dm^2 \frac{\sigma_1
(m^2)}{k^2-m^2+i\epsilon} \right]\ , \ee
so that we obtain
\be
\mbox{FT}\langle
\partial^{\lambda}F_{\lambda\mu},A_{\nu}\rangle =-R
\frac{k_{\mu}k_{\nu}}{k^2-i\epsilon} +(k^2g_{\mu\nu}-k_{\mu}k_{\nu})
\int dm^2 \frac{\sigma_1 (m^2)}{k^2-m^2+i\epsilon}\ .\ee
Then, thanks to the condition 2), for the color charge $Q$ we have
\be \mbox{FT}\langle g J_{\mu},A_{\nu}\rangle
=-(k^2g_{\mu\nu}-k_{\mu}k_{\nu}) \int dm^2 \frac{\sigma_2
(m^2)}{k^2-m^2+i\epsilon}\  .\ee
Finally we have
\be \mbox{FT}\langle i\delta \overline{\delta}A_{\mu},A_{\nu}\rangle
=-C \frac{k_{\mu}k_{\nu}}{k^2+i\epsilon}
-(k^2g_{\mu\nu}-k_{\mu}k_{\nu}) \int dm^2 \frac{\sigma_3
(m^2)}{k^2-m^2+i\epsilon }\ .\ee
Substituting these expressions for the three terms in
(\ref{green_funct}), we obtain a simple relation:
\be R=C\ .\ee
Here $R$ represents the residue of the massless pole of the gluon
propagator and $R=0$ would mean the absence of the massless gluon.
Thus the condition (\ref{conf_cond}) for color confinement implies
the gluons to be massive. On the other hand,when $C\neq 0$ we have
massless gluons, since $R\neq0$.

In this way we have found a connection between color confinement and
the gluon mass. We call our attention, however,  to the fact that
all the expressions given in this section are unrenormalized ones,
and we have to refine our arguments in the renormalized theory. This
is done in the next section.

\setcounter{equation}{0}
\section{Renormalization Group}

In order to justify the arguments developed in the preceding section
we rely on the RG. The generator of this group is given by
\be{\cal D} = \mu \frac{\partial}{\partial \mu} + \beta (g)
\frac{\partial}{\partial g} - 2 \alpha \gamma_{V} (g,\alpha)
\frac{\partial}{\partial \alpha}\ ,\ee
where $\mu$ denotes the renormalization point with dimension of mass
and $\gamma _{V}$ is the anomalous dimension of the color gauge
field.

We introduce a set of running parameters by
\bea\overline{g}(\rho)&=&  \exp(\rho {\cal D}) \cdot g , \cr
\overline{\alpha}(\rho)&=&  \exp(\rho {\cal D}) \cdot \alpha , \cr
\overline{\mu}(\rho)&=&  \exp(\rho {\cal D}) \cdot \mu =e^{\rho}\mu
\ ,\eea
where $\rho$ denotes the parameter of RG. The asymptotic limits of
these running parameters are denoted by
\be \bar{g} (\infty) = g_{\infty},\ \  \bar{\alpha} (\infty) =
\alpha_{\infty}\ . \ee
Asymptotic freedom characteristic of non-Abelian gauge theories is
expressed as
\be\label{asympt_freedom}g_{\infty} = 0\ . \ee
In solving the RG equations we introduce the ultraviolet cut-off so
as to avoid divergences in the unrenormalized expressions, then in
the asymptotic limit the running parameters and their functions
reduce to bare or unrenormalized ones.  The relationship
(\ref{asympt_freedom}), however, is obtained in the limit of the
infinit cut-off.

Take, for instance, a running parameter $\bar a(\rho)$, then in the
asymptotic limit we have:
\be\bar a(\infty)=a^{(0)}\ ,\ee
where $a^{(0)}$ denotes an unrenormalizable or bare expression and
this relationship gives a boundary condition for $\bar a(\rho)$ when
we solve the RG equation. On the other hand, the renormalized
expression $a$ is simply given by
\be\bar a(0)=a\ .\ee
Let us consider two multiplicatively renormalizable running
parameters $\bar a(\rho)$ and $\bar b(\rho)$ and assume that they
satisfy RG equations of the form
\bea\label{RGE} \frac{d}{d\rho}\bar a(\rho)=\gamma_a(\rho)\bar
a(\rho)\cr \frac{d}{d\rho}\bar b(\rho)=\gamma_b(\rho)\bar b(\rho)\ ,
 \eea
and the boundary condition
\be\label{bound_cond} a^{(0)}=b^{(0)}\ \ \mbox{or}\ \ \bar
a(\infty)=\bar b(\infty)\ ,\ee
then $\bar a(\rho)$ can be expressed in terms of $\bar b(\rho)$ as
\be\label{eq_param}\bar a(\rho)=\bar
b(\rho)+\int_{\rho}^{\infty}d\rho'[\bar\gamma_a(\rho')-\bar\gamma_b(\rho')]\bar
b(\rho')\cdot\exp\int_{\rho'}^\rho d\rho''\bar\gamma_a(\rho'')\ .\ee
The difference $a-b=\bar a(0)-\bar b(0)$ shows up in the
renormalized expressions as the coefficient of the so-called
Schwinger term \cite{Schwinger}. The above result was obtained by
combining (\ref{RGE}) and (\ref{bound_cond}), thereby keeping the
cut-off finite. We assume, however, that (\ref{eq_param}) is valid
also in the limit of an infinite cut-off and take it for granted in
the renormalized version. Then, when one of them vanishes
identically, for instance, $\bar b(\rho)=0$, we may conclude that
the other also vanishes, $\bar a(\rho)=0$.

In the preceding section we have derived an equality
\be R^{(0)}=C^{(0)}\ \ \mbox{or}\ \ \bar R(\infty)=\bar C(\infty)\
,\ee
and an application of the above arguments lead to the equivalence of
the following two statements:
\be\label{equiv} R=0\leftrightarrows C=0\ .\ee
In this connection it should be emphasized that (\ref{equiv}) was
derived by comparing the residues of the massless spin zero poles in
(\ref{green_funct}), whereas the equivalence of the two states
\be Z_3^{-1}=0\leftrightarrows C=0\ee
mentioned in Section 2 can be derived by replacing the T-products by
equal-time commutators in (\ref{green_funct}). In this way we have
justified the connection between color confinement and the gluon
mass.

In QED we have $C=1$ so that there is no charge confinement and the
photon is massless. This is a typical example to show the relevance
of the connection discussed in this paper.

Next we shall check when the confinement condition is satisfied and
for this purpose we consult the following identity
\cite{color_conf}-\cite{4}:
\be Z_3^{-1}   = \frac{\alpha}{\alpha_{\infty}}\ .\ee
The parameter $\alpha_\infty$ can assume the following three values:
\be \alpha_\infty=-\infty,\ 0,\ \alpha_0\ ,\ee
where
\be \alpha_{0} = \frac{1}{3} (13- \frac{4}{3} N_f)\ .\ee
$N_f$ denotes the number of quark flavors, and for simplicity we
shall confine ourselves to the simple case of positive $\alpha_{0}$,
then we have \cite{10,11'}
\be  Z_3^{-1}= \left \{\begin{tabular}{lll}
$0\ ,$ & for & $\alpha\leq0$ \\
$\alpha/\alpha_0\ ,$ &
for & $\alpha>0$\\
\end{tabular}
\right.\ee
corresponding to
\be  \alpha_\infty= \left \{\begin{tabular}{lll}
$-\infty\ ,$ & for & $\alpha<0$ \\
$0\ ,$ &
for & $\alpha=0$\\
$\alpha_0\ ,$ &
for & $\alpha>0$\\
\end{tabular}
\right.\ee

Thus we may conclude that color confinement is realized for
$\alpha\leq 0$ and the gluon turns out to be massive, whereas it is
not realized for $\alpha>0$ and the gluon remains massless. It might
occur to us that physical results should be independent of the
choice of the gauge parameter, but it is not always the case.

In showing the equivalence between theories of two different gauge
parameters, say $\alpha$ and $\alpha+\Delta\alpha$, we usually refer
to power series expansions of Green's functions in $\Delta\alpha$.
If this series converges, we may infer the gauge independence, but
eventually $\Delta\bar \alpha(\rho)$ diverges in the limit
$\rho\to\infty$. This is really the case when $\alpha>0$ and
$\alpha+\Delta\alpha<0$ or $\alpha<0$ and $\alpha+\Delta\alpha>0$.
Then we have to conclude that there are two distinctive phases in
QCD depending on the signature of the gauge parameter although it
contradicts our na\"ive belief. It should also be mentioned that the
transition between these two phases cannot take place since the
gauge parameter is not an adjustable dynamical variable and only the
confinement phase seems to be realized in nature.

\end{document}